# Impact of Vibrotactile Triggers on Mental Well-Being through ASMR Experience in VR

Danyang Peng[1][0000−0002−0425−9760], Tanner Person[1][0000−0002−9546−8903], Ximing Shen[1][0000−0003−4272−0068], Yun Suen Pai[2][0000−0002−6090−2837], Giulia Barbareschi[1][0000−0001−9036−3566], Shengyin Li[1][0009−0004−5290−982X], and Kouta Minamizawa[1][0000−0002−6303−5791]

[1] Keio University Graduate School of Media Design, Yokohama, Japan @kmd.keio.ac.jp
[2] University of Auckland, Auckland, New Zealand
yun.suen.pai@auckland.ac.nz

**Abstract.** Watching Autonomous Sensory Meridian Response (ASMR) videos is a popular approach to support mental well-being, as the triggered ASMR tingling sensation supports de-stressing and regulating emotions. Therefore, there is increasing research on how to efficiently trigger ASMR tingling sensation. Tactile sensation remains unexplored because current popular ASMR approaches focus on the visual and audio channels. In this study, we explored the impact of tactile feedback on triggering ASMR tingling sensation in a Virtual Reality (VR) environment. Through two experimental studies, we investigated the relaxation effect of a tactile-enabled ASMR experience, as well as the impact of vibrotactile triggers on the ASMR experience. Our results showed that vibrotactile feedback is effective in increasing the likelihood of ASMR tingling sensation and enhancing the feeling of comfort, relaxation, and enjoyment.

## 1 Introduction

Autonomous Sensory Meridian Response (ASMR) is a phenomenon that occurs when individuals experience tingling sensations in response to specific audio or visual stimuli. It is a relaxing sensory experience [7] and has been proven effective in reducing stress, inducing relaxation, aiding sleep, and even relieving pain symptoms [1, 7, 21]. Typically presented through online videos or audio recordings, the growing ASMR community and increasing proliferation of online content contribute to its continuous exploration and dissemination. Despite its popularity, the potential significance of tactile triggers in ASMR has been, to date, often overlooked. Touch sense serves not only as a concurrent synesthetic experience in the context of ASMR but also functions as a potent trigger [9]. To further understand the potential of vibrotactile triggers in ASMR, we designed our study guided by the following research questions:

**RQ1:** Does a VR-based ASMR experience have an impact on mental wellness?

**RQ2:** Is the vibrotactile feedback that has an impact on the ASMR tingling sensation, and the mental wellness effect?

We designed a VR-based remote multiplayer real-time interaction system and connected wearable devices with vibrotactile feedback, creating multi-modal ASMR triggers for visual, auditory, and tactile channels. We then conducted two studies. Study 1 was a quantitative study designed to understand the effect of the multimodal ASMR experience on user mental well-being using the Experienced Pleasantness of Touch Questionnaire(EPTQ) [10]. Study 2 was a quantitative study designed to understand the impact of vibrotactile triggers on the ASMR experience using the EPTQ, ASMR-15 scale [16], and the Avatar Embodiment Questionnaire(AEQ) [18]. We found that the multimodal triggers ASMR experience has a positive effect on users' wellness, and the vibrotactile triggers are effective in eliciting ASMR tingling sensation.

Here, we summarize our main contributions:

(1) Designed a multi-modal trigger ASMR experience based on VR;

(2) Explored the impact of vibrotactile triggers in VR-based ASMR experience, and provided design suggestions for vibrotactile wearable devices to enhance the ASMR experience;

(3) Evaluated the impact of multimodal triggers ASMR experience on mental well-being.

## 2 Related Works

### 2.1 ASMR for Mental Wellbeing

In recent years, ASMR videos have offered a private, and non-contact form of entertainment. The growing ASMR community has motivated creators to explore innovative triggers [17, 2, 11]. With the emergence of various technologies. Some creators have employed diverse technological interactive methods to enhance users' sensory experiences through the utilization of technology. For instance, ASMR-inspired soundwave toolkits serve to encourage users to explore their surroundings and cultivate personal, intimate attention practices. It demonstrates that ASMR media can convey relaxation and reflective design in a meaningful manner [12]. The use of multimodal triggers helps users improve their ASMR experience [19].

The triggering factors of ASMR often involve multiple sensory modalities. Whispering, personal attention, and crisp sounds, slow movements were reported to induce the most tingling sensation in common triggers [1, 15]. Niu et al. [17] analyzed 2,663 ASMR-related videos on YouTube, and summarized three main features of ASMR videos on YouTube: diverse social connections, a relaxing sense of bodily intimacy, and sensory-rich observational activities. Poerio et al. [20] found that physical contact with the body emerged as the most widely acknowledged triggering factor, with the highest intensity. Additionally, ASMR has been described as an exaggerated form of pleasurable responses to interpersonal tactile interactions, such as hugs or emotional expressions through the body [9, 27].

### 2.2 VR for Mental Well-Being

Virtual Reality technology has immense potential for applications in emotional regulation. Roy et al. designed four virtual environments to assist patients with social anxiety disorder in dealing with anxiety has confirmed that virtual reality therapy can effectively treat social anxiety disorder [22]. VR technology can enhance presence and provide an engaging environment for creating meditation experiences, and therefore users can engage in meditation practices within it [13, 23, 24]. Virtual Restorative Environments (VRE) highlight VR technology's unique capacity for emotional healing through immersive simulated experiences [26]. VR has been proven effective in emotional regulation, highlighting its potential to deliver personalized therapy and enhance mental health through immersive, calming experiences.

### 2.3 Vibrotactile Stimulation for Mental Well-Being

Vibrotactile stimulation has been widely used in fields such as entertainment and medicine to support mental health and well-being [14, 24, 3]. Auditory and tactile devices have been leveraged to assist users with falling asleep and relaxing [5]. At the same time, when used in conjunction with VR, tactile devices can enhance therapeutic effects, making them applicable for treating mental illnesses, including schizophrenia [4]. Vibrotactile devices may compensate for the shortcomings of solely relying on synesthetic touch in traditional ASMR experiences, to help users better regulate emotions and relieve stress. The role of tactile feedback devices in ASMR experiences is worth exploring.

## 3 Designing asmVR

### 3.1 Workshop Co-design

To enhance the design of the ASMR experience, we invited 7 ASMR enthusiasts(5 females, ages between 20-30). Participants did not only have extensive ASMR experiences but also used ASMR as a routine recreational method for stress relief. Participants engaged in a structured one-hour-long discussion on an online videoconferencing system, which incorporated hands-on activities on the Miro collaborative platform, the topics of the discussion centered around user preferences and the designers' prior experiences of ASMR.

The structure of the workshop could be summarised as follows. First, the physiological principles of ASMR and its current research status were introduced. Subsequently, three ASMR videos were played to facilitate participants in revisiting the tingling sensations associated with ASMR. Progressing through three stages, namely trigger design, scenario design, and character design, the workshop established a face-to-face interactive format primarily centered around touch and whispering. To cater to individual preferences, distinctions were made among ASMRtist's avatars, involving male, female, and cartoon animal categories[3]. Simultaneously, the virtual avatar roles for users were positioned as neutral, contributing to the design foundation of this study.

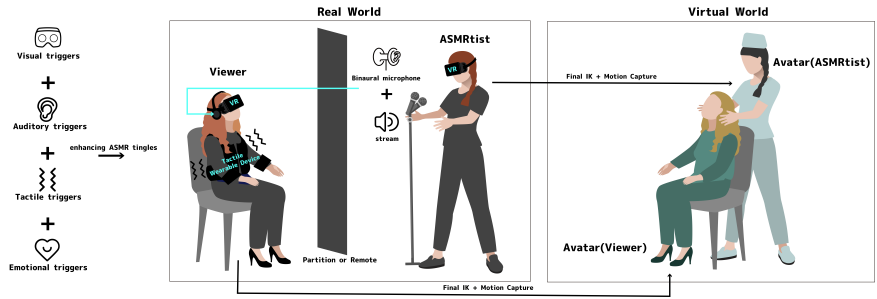

**Fig. 1.** asmVR system concept

### 3.2 System Description

Based on the co-design workshop and existing research, this study aims to enhance ASMR experiences for users by building a multiplayer VR system based on Unity3D[4]. It considers four major triggers of ASMR: (1) Visual: The visual component primarily employs VR-based, multiplayer face-to-face interaction, where ASMRtists and participants possess their respective virtual avatars. Motion capture by Quest Pro and IK enables users to control their virtual characters. (2) Auditory: The auditory component encompasses real-time or pre-recorded ASMRtist whispers during interactions, as well as touch sounds produced during touches. All sounds are captured or recorded using KU100 binaural microphone[5], and participants are required to wear headphones throughout the experience. (3) Tactile: Tactile sensation aspects include virtual touch sensations and vibration feedback. We chose the Bhaptic Vest, arms, and gloves for vibration feedback[6](with eccentric rotating mass actuators operating at the vibration frequency of approximately 90 Hz). (4) Emotional/interpersonal: Emotional aspects are conveyed through ASMRtists' whispers and slow touch(1-10 cm/s) during the interactive process, as well as the virtual avatar of the ASMRtist. By Fig. 2(d) when viewed from the perspective of an ASMRtist, tactile sensors are visualized as gray spheres in VR. The ASMRtist and the avatar of the audience are seated face-to-face within the VR environment, ensuring a spatial arrangement wherein the ASMRtist can touch all the gray spheres distributed across the audience's body. Each gray sphere on the audience's avatar corresponds to a sensor located at the respective position on the body. Upon the ASMRtist's interaction with a gray sphere, the corresponding vibrational sensor activates, aligning tactile contact locations with vibrational feedback positions. Concurrently, pre-recorded 3D binaural friction sounds are played. To enhance the immersive experience for the audience, the gray spheres are intentionally concealed from the audience's viewpoint, as illustrated in Fig. 2(c).

---

[3] https://hub.vroid.com/en/models

[4] https://unity.com/

[5] https://www.neumann.com/en-en/products/microphones/ku-100/

[6] https://www.bhaptics.com/

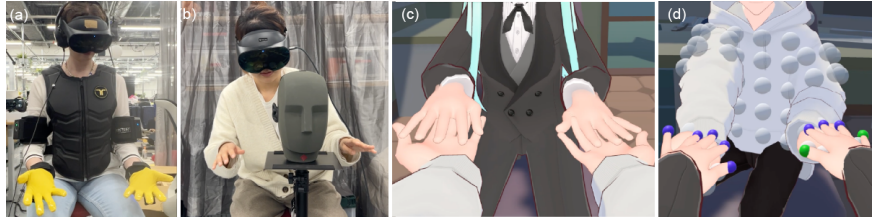

**Fig. 2.** (a)Participant wearing vibrotactile feedback devices. (b)Remote ASMRtist is performing in real time. (c)Viewer's Perspectives. (d)ASMRtist's Perspectives.

This research investigates the potential of the proposed system in alleviating the escalating psychological stress and anxiety faced by individuals. The objective is to use multi-modal triggers, aiming to provide users with a novel and immersive ASMR experience, with the ultimate goal of assisting users in stress relief and emotional regulation.

## 4 Study 1

### 4.1 Method and Procedure

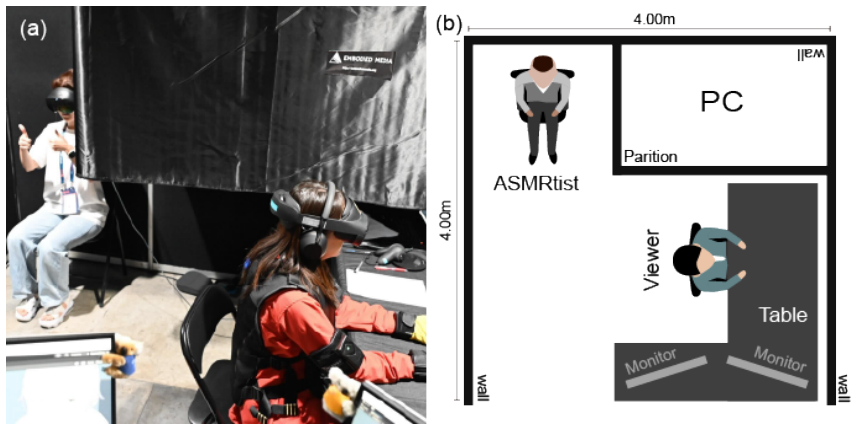

**Fig. 3.** (a) Demonstration process. (b) Floor plan of the booth

To evaluate whether the proposed method could have an impact on the mental health of users, we conducted user testing at an exhibition [19]. Participants visiting our booth(as Fig. 3(a)) were proactively approached and queried regarding their willingness to assist in completing a survey questionnaire to evaluate the impact of ASMR on mental well-being. If participants agreed, they were asked to sign a consent form; otherwise, they were invited solely to experience a demonstration.

We designed a demo lasting about 3-4 minutes based on real-time two-person VR interaction, with pre-trained personnel posing as remote ASMRtist, and randomly recruited participants as viewers. Users needed to wear gloves, arms, vests that provide vibrotactile feedback, Quest Pro VR headsets, and a noise-canceling headphone and sit in a chair. In the initial stage, users were asked to select their preferred ASMRtist virtual avatars. Available options included three distinct ASMRtist avatars: a female avatar, a male avatar, and a cartoon fox avatar. This was done to proactively mitigate potential discomfort that certain users may experience due to perceived proximity in social interactions with individuals of the same or opposite gender. Notably, the female and cartoon fox avatars employed female whispering for auditory stimulation, while the male avatar used a male voice but with the same dialogue and time.

The demo used recorded binaural audio, including whispers such as "Please relax", "Could you feel me touch your hands?", "Now I want to start touching your back", etc., as both auditory triggers for the audience and motion prompts for ASMRtist. ASMRtist sequentially touched users' hands, wrists, arms, front and back of the body, as well as hair and cheeks in VR, ensuring consistency in each demonstration process.

A purposefully designed survey was created to gather participant age and gender information. ASMR is considered to share common psychological and biological mechanisms with affective touch, described as a means to convey emotions through touch, enhancing interpersonal communication [6]. ASMR achieves physiological regulation by eliciting specific responses to emotional stimuli, such as caressing and whispering, through different sensory systems [25]. Therefore, this study selected the **Experienced Pleasantness of Touch Questionnaire(EPTQ)** [10]. This questionnaire comprises a total of seven items, encompassing five positive emotions and two negative emotions. Participants were required to rate their subjective experiences for each item on a Numerical Rating Scale (NRS) ranging from 0 (not at all) to 10 (very much). Participants were instructed to complete the survey both before and after engaging in the experiential demonstration.

### 4.2 Participants

The present experiment recruited 58 participants through random selection. Among them, 3 individuals were excluded from the analysis due to equipment connectivity issues, rendering their data deemed invalid. 55 participants($M = 34, F = 18, Other = 3$) completed all experimental sessions. In terms of age distribution($Mean = 31.22, SD = 9.60$), 5 participants opted not to reveal their age. It is noteworthy that this experiment did not specifically target ASMR responders(people who can feel the tingling sensation of ASMR).

### 4.3 Result

The results of study 1, as depicted in Fig. 4, the experimental data obtained have all passed the Normality test. The T-test was conducted with a 95% confidence interval to assess the contrast between before and after groups for participants regarding their pleasantness changes.

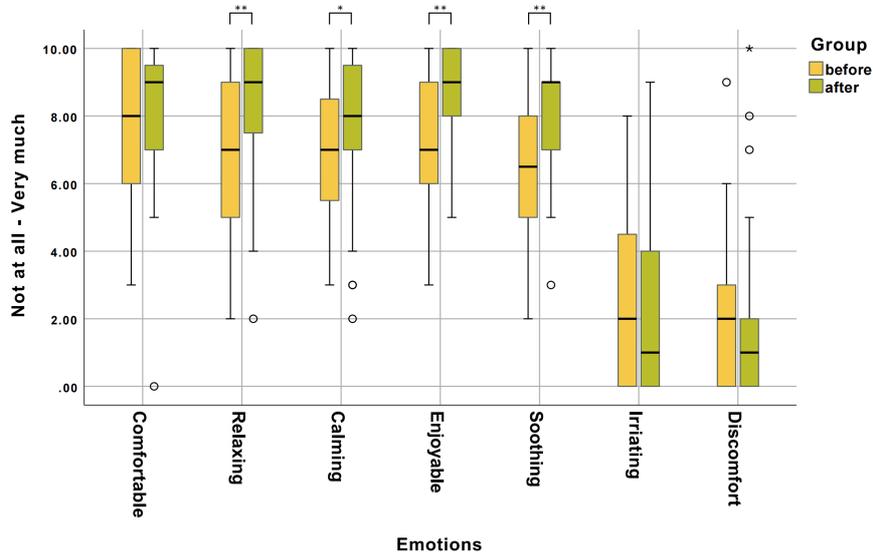

**Fig. 4.** The result of EPTQ in study 1

The results revealed no significant change in comfort level, as evidenced by $t(54) = -1.74, p = 0.087$. In contrast, the "Relaxing" dimension exhibits a highly statistically significant decrease with $t(54) = -4.20, p < 0.001$. Conversely, the "Calming" sub-scale shows statistically significant change with $t(54) = -2.50, p = 0.015$. The "Soothing" sub-scale demonstrates a highly statistically significant decrease with $t(53) = -5.42, p < 0.001$, while the "Enjoyable" condition also exhibits a highly statistically significant decrease with $t(54) = -6.09, p < 0.001$. The "Irritability" sub-scale shows no significant change, as indicated by $t(54) = 0.90, p = 0.371$, and the "Discomfort" one similarly exhibits no significant change with $t(54) = 1.12, p = 0.266$.

## 5 Study 2

### 5.1 Method and Procedure

The goal of Study 2 was to assess the influence of tactile feedback on users in VR-based ASMR experiences. The experience, devices, and procedures were kept consistent with the details outlined in Study 1.

Within-group measures were used to evaluate the impact of vibrotactile feedback. In the initial phase, the first group of participants engaged in the VR-based ASMR experience incorporating vibrotactile feedback. Subsequently, they completed questionnaires.

After a 30-minute interval, participants in this group then underwent the VR-based ASMR experience without vibrotactile feedback, after which they completed the same questionnaires and participated in a one-on-one interview. To prevent bias, the order of experiences was inverted for the second group.

About questionnaires, this experiment employed the **EPTQ**, as described in Study 1, alongside two additional scales.

The **ASMR-15 scale** is commonly utilized for assessing individuals' proclivity toward ASMR [16]. This scale encompasses a total of 15 questions distributed across four altered dimensions: consciousness, sensation, relaxation and affect. Each question is assigned a score ranging from 1 to 5. The ASMR total score is derived by summing the scores obtained from the questions within the four dimensions. A higher score on the ASMR-15 scale indicates a greater inclination towards ASMR in the user.

In the VR environment, participants experienced the avatar from the first perspective, feeling that a self-incarnation has replaced their body, and the new body is the source of sensation. To evaluate the impact of vibration tactile triggers on the elicitation of the embodiment illusion, we utilized the **Avatar Embodiment Questionnaire(AEQ)** [18], participants were asked to rate their subjective experiences for each item on the NRS ranging from 1 (strongly disagree) to 7 (strongly agree). A higher score on the embodiment score indicates that the user experiences a stronger embodiment illusion in their virtual avatar.

### 5.2  Participants

20 participants($M = 10, F = 10, Mean = 25.05, SD = 2.61$) were recruited through random selection. Before the start of the experiment, all participants were provided with comprehensive information regarding the experimental procedures. Participants will only proceed with the experiment after signing the consent form for the experiment. Participants were stratified into two groups, each comprising 5 males and 5 females.

Overall, 13 individuals self-reported as ASMR responders, indicating their ability to perceive ASMR tingling sensations or having previously experienced such sensations. Conversely, 7 participants identified as non-ASMR responders, expressing an inability to perceive ASMR tingling sensations or a lack of prior experiences with similar sensations.

### 5.3  Result

This study used SPSS software to analyze data, which are shown in Fig. 5, the experimental data obtained have all passed the Normality test and conform to a normal distribution. The T-test was conducted with a 95% confidence interval to assess the contrast between the with vibrotactile device group and the without vibrotactile device group for participants. For the sensation of "Comfortable", with vibrotactile devices ($Mean = 7.65, SD = 2.23$) compared to without the devices ($Mean = 6.35, SD = 2.21$), the difference shows a statistically significant change, $t(19) = 2.46, p = 0.024$. "Relaxing", with vibrotactile devices ($Mean = 8.20, SD = 1.20$), without the devices ($Mean = 6.75, SD = 2.44$), the difference shows statistically significant change, $t(19) = 2.96, p = 0.008$. For "Calming", although the mean score with vibrotactile devices ($Mean = 7.50, SD = 1.73$) is higher than without the devices ($Mean =$

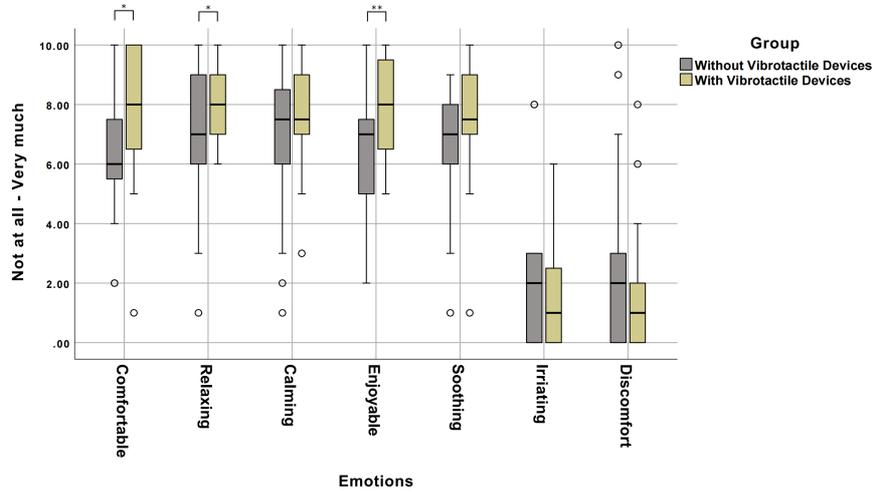

**Fig. 5.** The result of EPTQ in study 2

$6.95, SD = 2.50$), the difference is not statistically significant, $t(19) = 0.94, p = 0.357$. Similarly, for "Soothing", the difference between with vibrotactile devices ($Mean = 7.50, SD = 1.99$) compared to without the devices ($Mean = 6.65, SD = 1.98$), is not statistically significant, $t(19) = 1.39, p = 0.181$. For "Enjoyable", the mean score is higher with vibrotactile devices ($Mean = 8.00, SD = 1.59$) compared to without the devices ($Mean = 6.40, SD = 1.90$), and the difference is highly statistically significant, $t(19) = 4.21, p < 0.001$. Finally, for "Irritating" ($p = 0.076$) and "Discomfort"($p = 0.085$), the difference shows no statistically significant change.

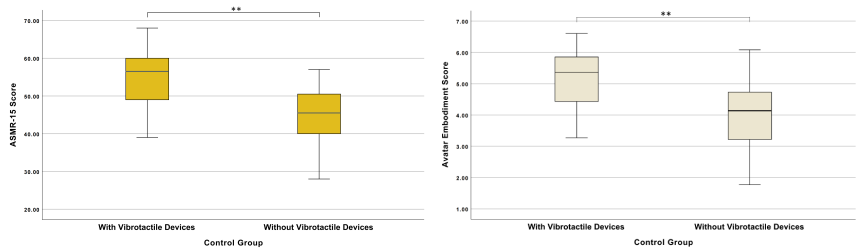

**Fig. 6.** (left)the result of ASMR-15. (right) the result of AEQ

When assessing the likelihood of ASMR experiences Fig. 6(left), the mean score was higher with the vibrotactile devicest($Mean = 54.55, SD = 8.29$) than without($Mean = 45.65, SD = 7.86$), and the difference was found to be highly statistically significant, $t(19) = 4.07, p < 0.001$. Moreover, the level of avatar embodiment as Fig. 6(right), compared between with devices group($Mean = 5.13, SD = 0.99$), and without device group($Mean = 4.00, SD = 1.10$), was shown to be statistically significant, $t(19) = 4.66, p < 0.001$.

## 6  Discussion

In study 1, the findings of our experiment demonstrate that a VR-based multi-modal trigger ASMR experience can assist users in relaxation, calming, soothing, and enjoyment. However, no significant improvement was observed to self-reported irritability or discomfort. This study points to the ability of our system to support certain aspects of stress relief and emotional regulation [1, 21, 9]. Interestingly, during the exhibition process, some participants who had never experienced ASMR reported a tingling sensation at the back of the head or the back region during oral interviews, and some participants also reported a vibration feedback sensation in the head, indicating vibration hallucinations, as devices were only worn on hands, arms, and body.

In study 2, results showed that vibrotactile feedback enhanced user perceptions of comfort, relaxation, and enjoyment, with positive emotional impacts being more pronounced than negative ones. This aligns with research showing how vibrotactile feedback had been leveraged in a variety of contexts from entertainment, medical and other fields, to helping users regulate their emotions [25]. Users utilizing vibrotactile feedback devices have also shown a significant increase in propensity for ASMR. Concurrently, there is a significant improvement in the scores for embodiment concerning one's avatar within VR. This indicates that vibrotactile feedback may augment the immersion of users in the virtual environment by facilitating a more convincing self-avatar embodied. This has also been corroborated in the research conducted by Jakob Fröhner, which demonstrates that haptic feedback facilitates user immersion in virtual reality environments [8].

Overall, this results show that vibrotactile feedback could heighten the realism perceived during virtual touch interactions and overall immersion in the virtual realm, thereby leading to an elevated inclination towards ASMR. However, it is noteworthy that some participants reported that overly strong vibration tactile sensation could lead to arousal during the experience, suggesting that in VR-based ASMR experiences, participants may prefer a gentle and linear vibration.

## 7  Conclusions

This study investigates the effect of vibrotactile triggers on a VR-based ASMR experience. The results suggest that vibrotactile triggers can enhance ASMR tendencies and the embodied feeling with virtual avatars, which supplemented the impact of vibrotactile triggers on the ASMR experience process. Comfort was highest for back and hands areas, with linear rather than localized vibrations, preferred for a more realistic touch sensation. This study underscores the potential of VR-based remote ASMR experiences for mental well-being and reveals the possibility of using vibration tactile triggers in the field of psychotherapy in the future.

**Acknowledgments.** This work is supported by JST Moonshot R&D Program "Cybernetic being" Project (Grant number JPMJMS2013) and JST COI-NEXT "Minds1020Lab" Project Grant Number JPMJPF2203.

**Disclosure of Interests.** The authors have no competing interests to declare that are relevant to the content of this article.